\documentclass[sigconf]{acmart}

\AtBeginDocument{%
  }

\usepackage{algorithmic}
\usepackage{graphicx}
\usepackage{textcomp}
\usepackage{array, booktabs, multirow, tabularx, xcolor, longtable, threeparttable}
\usepackage{makecell}
\usepackage{float} 
\usepackage{tcolorbox}

\usepackage{pifont}

\usepackage{makecell}

\newif\ifshowchanges
\showchangestrue 

\ifshowchanges
  \makeatletter
  \let\comment\@undefined
  \let\endcomment\@undefined
  \makeatother
  \usepackage[authormarkup=none]{changes}
\else
  \makeatletter
  \let\comment\@undefined
  \let\endcomment\@undefined
  \makeatother
  \usepackage[final]{changes}
\fi

\setcopyright{acmlicensed}
\copyrightyear{2018}
\acmYear{2018}
\acmDOI{XXXXXXX.XXXXXXX}
\acmConference[Conference acronym 'XX]{Make sure to enter the correct
  conference title from your rights confirmation email}{June 03--05,
  2018}{Woodstock, NY}
\acmISBN{978-1-4503-XXXX-X/2018/06}

\begin{document}

\title{
Exploring and Complementing End-Users’ Requirements in IoT-enabled Systems
}

\author{Haotian Li}
\authornotemark[1]
\email{51275902042@stu.ecnu.edu.cn}
\affiliation{%
  \institution{East China Normal University}
  \city{Shanghai}
  \country{China}
}

\author{Xiaohong Chen}
\authornotemark[1]
\email{xhchen@sei.ecnu.edu.cn}
\affiliation{%
  \institution{East China Normal University}
  \city{Shanghai}
  \country{China}
}

\author{Zhi Jin}
\authornotemark[2]
\email{zhijin@whu.edu.cn}
\affiliation{%
  \institution{Wuhan University}
  \city{Wuhan}
  \country{China}
}

\author{Shuyuan Xiao}
\authornotemark[1]
\email{10235101446@stu.ecnu.edu.cn}
\affiliation{%
\institution{East China Normal University}
\city{Shanghai}
\country{China}
}
\author{Chenxu Wang}
\email{1160535294@qq.com}
\affiliation{%
\institution{Shanghai DianJi University}
\city{Shanghai}
\country{China}
}
\author{Haoxiang Yan}
\authornotemark[1]
\email{10225101531@stu.ecnu.edu.cn}
\affiliation{%
\institution{East China Normal University}
\city{Shanghai}
\country{China}
}
\author{Xiaoyi Chen}
\authornotemark[1]
\email{10220740414@stu.ecnu.edu.cn}
\affiliation{%
\institution{East China Normal University}
\city{Shanghai}
\country{China}
}

\begin{abstract}

 End users create IoT automation rules via trigger-action programming, but their expressions are often fragmented—capturing device operations rather than high-level intents. This gap leads to missing conditions, logical conflicts, and overlooked safety constraints, risking hazardous behaviors. 
To address this, we propose an intent-driven requirements completion approach that reframes rule completion as a dual process: reconstructing intent from fragmented rules, then regenerating rules from that intent—with safety embedded throughout. We introduce a \textbf{Bidirectional Requirements Traceability Tree}, a three-layer model linking rules, intents, and quality concerns, and design a multi-agent framework that combines LLM reasoning with structured traceability. This enables completions that are both functionally complete and inherently safe, while remaining traceable and explainable.
Evaluation shows our method significantly outperforms the baselines, improving the rule completion rate by 43\% and reducing logical conflicts by over 21\%.
By grounding completion in intent understanding, we shift the paradigm from user to system responsibility, and from functional correctness to holistic trustworthiness.

\end{abstract}




\renewcommand{\shortauthors}{Anonymous Author, et al.}


\maketitle

\section{Introduction}

The Internet of Things (IoT) has transformed our physical environments from passive spaces into programmable ecosystems~\cite{taivalsaari2017roadmap}. Through Trigger-Action Programming (TAP) platforms such as IF-TTT~\cite{ifttt_platform}, Zapier~\cite{zapier_platform}, and smart home applications, end-users can now create custom automation rules-``if the temperature drops below 23$^oC$, turn on the heater''—that shape their surroundings to match their preferences. This democratization of automation empowers users to personalize their living and working environments without requiring professional programming expertise~\cite{ur2014practical, desolda2017empowering}.

However, while TAP rules are the primary vehicle through which users express their automation needs, they are fundamentally different from users' underlying intents~\cite{bian2021approach,chen2025expressing, corno2021users, wilson2015smart, cimino2025iot}. User intents are typically high-level goals they wish to achieve—such as ``keep the room comfortable''—which are inherently device-agnostic~\cite{chen2025expressing, mennicken2014today, davidoff2006principles, ghiani2017personalization}. In contrast, TAP rules are essentially device scheduling specifications that map specific trigger conditions to particular actions~\cite{chen2025expressing, zhang2019autotap}. This distinction gives rise to a critical challenge: end-user expressions are often fragmented and incomplete, resulting in quality issues such as logical conflicts or missing conditions in the generated TAP rules~\cite{zowghi2003interplay, jeong2025device}. In particular, non-functional quality concerns—most notably safety~\cite{curumsing2019emotion, brink2013addressing} and security~\cite{alqassem2014privacy, barbosa2019if, psychoula2018users}—are frequently overlooked when users create rules, and their absence can lead to unexpected or even hazardous system behavior~\cite{chen2021iotranx, funk2018addressing}. Therefore, the core difficulty lies in bridging the gap between users' partial expressions and the need for complete, consistent, and safe automation logic. 

To address this challenge, existing research has evolved from traditional methods to agent-based approaches. Early studies frame-d rule completion as a recommendation or association problem, using collaborative filtering~\cite{dominguez2018mashup} (e.g., SVD++) or clustering-based mining~\cite{liu2022device} to capture user–rule co-occurrence patterns. However, these methods were constrained by data sparsity and lacked deep semantic understanding. Subsequent deep learning based approach-es~\cite{cimino2025iot} attempted to map user needs into latent spaces via representation learning to enhance semantic comprehension, yet their performance depended heavily on annotation quality and reasoning capabilities remained limited. Current large language models (LLMs)~\cite{gao2024chatiot} leverage strong contextual understanding to generate rules, but they suffer from hallucinations—producing plausible yet risky rules that are unacceptable in physical deployments. The recent multi-agent methods~\cite{hong2023metagpt} introduce role collaboration and reflection mechanisms to decompose complex tasks, showing promise in general domains. However, when applied to TAP rule completion, their planning still relies on blind reasoning with pretrained knowledge, lacking interpretable theoretical foundations and leaving the completion process an uncontrollable black box.

We observe that although the TAP rules written by users are fragmented, they do not emerge randomly—each rule is a ``trace'' left by users attempting to realize some high-level intent. In other words, rules are the external manifestation of intents, while intents are the internal basis for rules. This led us to realize that the essential task of TAP rule completion is not to perform ``gap-filling'' at the sentence level, but rather to conduct ``reconstruction and restoration'' at the intent level. If we do not understand the intent behind a user's actions, any completion can only be blind speculation—and this is precisely the fundamental dilemma of existing completion approaches: they attempt to ``fill in rules'' without understanding the underlying user intent, resulting either in incomplete completions or in rules that seem plausible but are actually dangerous.

Therefore, we argue that a truly viable path must be ``\textbf{intent-driven}'': first understand what the user wants to do, then determine how to complete the rules. This requires reframing the completion problem as a bidirectional traceability task—both reconstructing high-level intents from fragmented rules and forward-deriving complete rule sets from those intents.
To achieve this, we propose the \textbf{Bidirectional Requirements Traceability Tree (BRT)}, a three-layer tree structure: the top layer represents high-level intents (e.g., ``maintain comfort''), the middle layer represents logical plans, and the bottom layer represents concrete TAP rules. It explicitly captures the ``one-to-many'' and ``many-to-one'' relationships between intents and rules, serving both as a mapping bridge and a completion backbone—inferring latent intents bottom-up and generating complete rules top-down, making the entire completion process traceable and explainable.
We design a multi-agent collaboration framework that integrates the reasoning capabilities of LLMs with structured traceability requirements.

We conduct experimental evaluations on both real-world and synthetic datasets. 
The results demonstrate that our method significantly outperforms the baselines, improving the rule completion rate by an absolute 43\% and reducing logical conflicts by over absolute 21\%.
However, the value of this work extends beyond algorithmic improvement: through the \textbf{BRT}, we elevate requirement completion from ``mechanical rule filling'' to ``traceable intent reconstruction,'' shifting the system from "passively executing instructions" to ``actively understanding intents.'' This transformation represents a critical step forward in the human-computer interaction paradigm—from ``users must specify clearly'' to ``systems strive to understand,'' and from ``functional implementation'' to ``trustworthy assurance.'' The main contributions of this paper are as follows:
\begin{itemize}
    \item We present \textbf{BRT}— A multi-layer requirements representation model bridging intents and rules. We introduce the concept of ``traceability'' from requirements engineering into the end-user programming domain for the first time, providing a theoretical foundation for understanding ``why users write rules the way they do.''
    \item  We design a multi-agent collaboration approach for intent-driven requirements completion in IoT. We reframe requirements completion as the construction and maintenance of \textbf{BRTs}, and design a multi-agent collaboration framework to realize this process, ensuring that every completion is traceable, explainable, and satisfies quality constraints.
    \item  We construct and open-source the first benchmark dataset for quality-aware requirements completion, featuring explicit annotations of users' implicit quality needs. This fills a critical gap in evaluation tools for this field and provides a reproducible comparison platform for future research.
\end{itemize} 
\vspace{-0.2cm}
\section{Framework and BRT}
This section provides an overview of our method, presents the collaborative control mechanism of the multi-agent framework, and defines the core requirements representation model \textbf{BRT}. 

\subsection{Overview}
We propose an intent-driven framework for TAP rule completion via bidirectional traceability. This framework is aligned with the bidirectional process: bottom-up reconstruction of high-level user intentions from fragmented rules, and top-down generation of comprehensive rule sets from these derived intents. Crucially, quality concerns are intrinsically embedded throughout this process. 

\begin{figure*}
    \centering
    \includegraphics[width=1\linewidth]{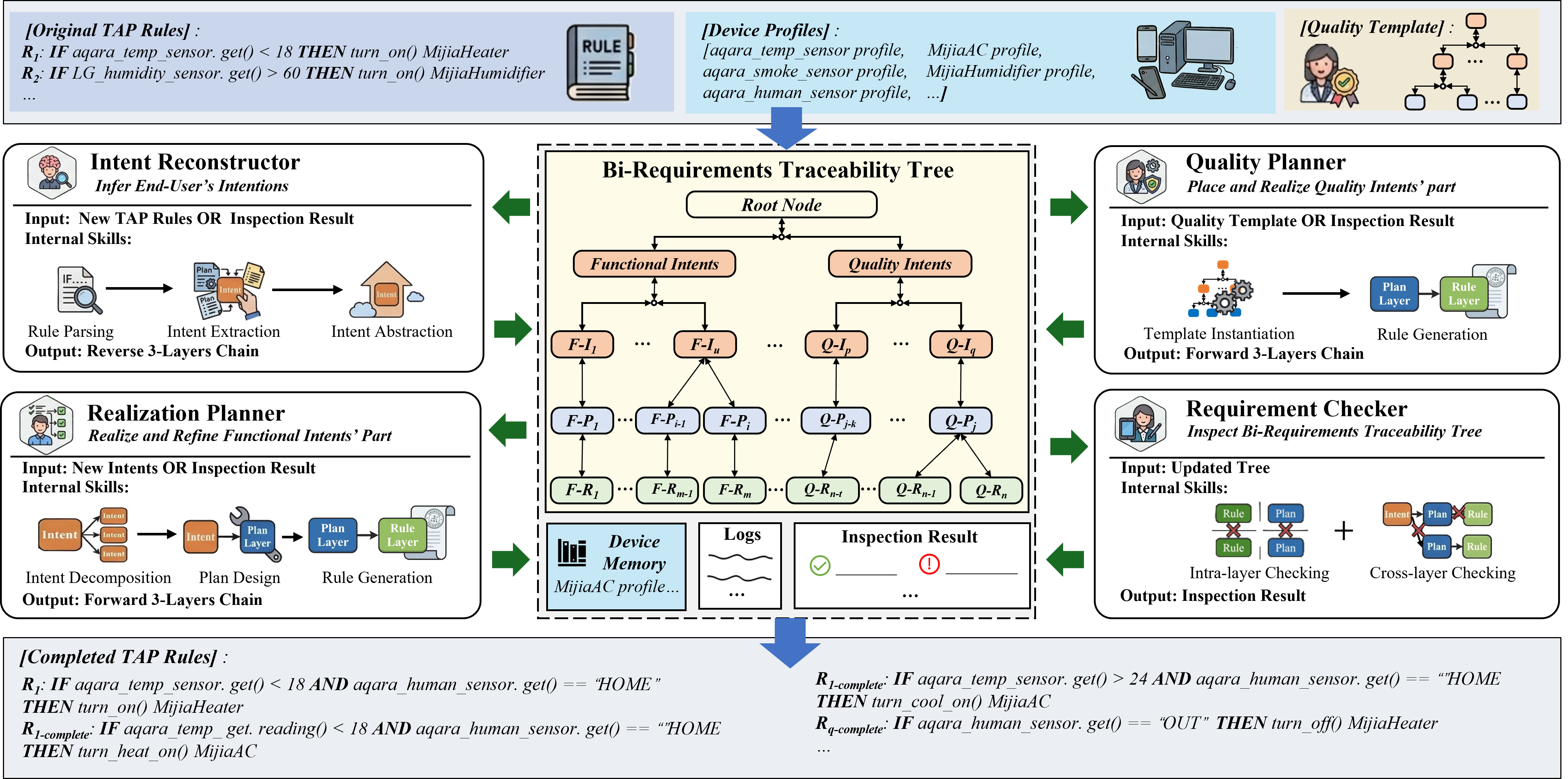}
    \caption{Overall Framework for Multi-agent Collaboration Approach Based on Bidirectional Traceability}
    \label{fig:framework}
\end{figure*}

The architecture comprises two primary components. First, the \textbf{BRT} provides a structured representation that explicitly models the hierarchical relationships among user intents, logical plans, and executable TAP rules, enabling traceable and explainable reasoning. Second, a synergistic multi-agent system as illustrated in Figure ~\ref{fig:framework} operations the \textbf{BRT} to overcome the ``black-box'' limitations of conventional generation methods. 

This system deploys four specialized agents with distinct responsibilities. \textit{Intent Reconstructor} infers missing intents and plans from fragmented inputs to capture underlying user goals. Concurrently, \textit{Quality Planner} integrates environment-specific quality requirements, while \textit{Realization Planner} translates the established intents into concrete, executable TAP rules. Finally, \textit{Requirement Checker} rigorously inspects the \textbf{BRT}. These agents collaborate via a closed-loop, multi-round iterative mechanism. If \textit{Requirement Checker} detects errors, it propagates targeted feedback to the relevant agents. Based on this feedback, \textit{Intent Reconstructor} revises intent nodes, \textit{Quality Planner} adjusts its integration strategy, and \textit{Realization Planner} optimizes the functional rules. This self-correction cycle repeats until the system passes validation, ultimately outputting a fully reliable and safe automation rule set.

\subsection{Collaborative Control mechanism}

\textbf{State Management Based on Shared Workspace:} Four agents collaborate with each other in a shared workspace that maintains the global state of the \textbf{BRT}. To preserve structural integrity, the workspace enforces strict access controls and fine-grained locking, permitting concurrent reads while mandating mutually exclusive writes.
Each agent possesses specific read/write privileges: 
\textit{Intent Reconstructor} infers and writes bottom-up intent and plan nodes from the user's fragmented rules; 
\textit{Quality Planner} instantiates and appends quality-oriented nodes based on environmental context; 
and \textit{Realization Planner} top-down expands intents into concrete plan and rule nodes. 
Conversely, \textit{Requirement Checker} operates with read-only access to the \textbf{BRT}, exclusively writing inspection result to the shared workspace. 
Furthermore, the workspace incorporates a device memory to maintain real-time information of available devices within the current environment, alongside an operation history log records every modification, thereby ensuring full traceability throughout the completion process.

\textbf{Deadlock Prevention and Exception Handling:} 
To prevent deadlocks, the framework employs a priority-based scheduling strategy. By assigning the highest priority to \textit{Intent Reconstructor}, the process mandates that foundational intents are established first. Following this initial construction, \textit{Quality Planner} and \textit{Realization Planner} enter a concurrent execution phase to independently synthesize the \textbf{BRT}’s quality and functional parts.
Exception handling protocols are embedded within the workspace to prevent cascading failures. For instance, if an agent generates structurally invalid components that compromise \textbf{BRT} integrity, the workspace outright rejects the write operation and issues diagnostic feedback to enforce immediate self-correction.

\begin{figure}
    \centering
    \includegraphics[width=1\linewidth]{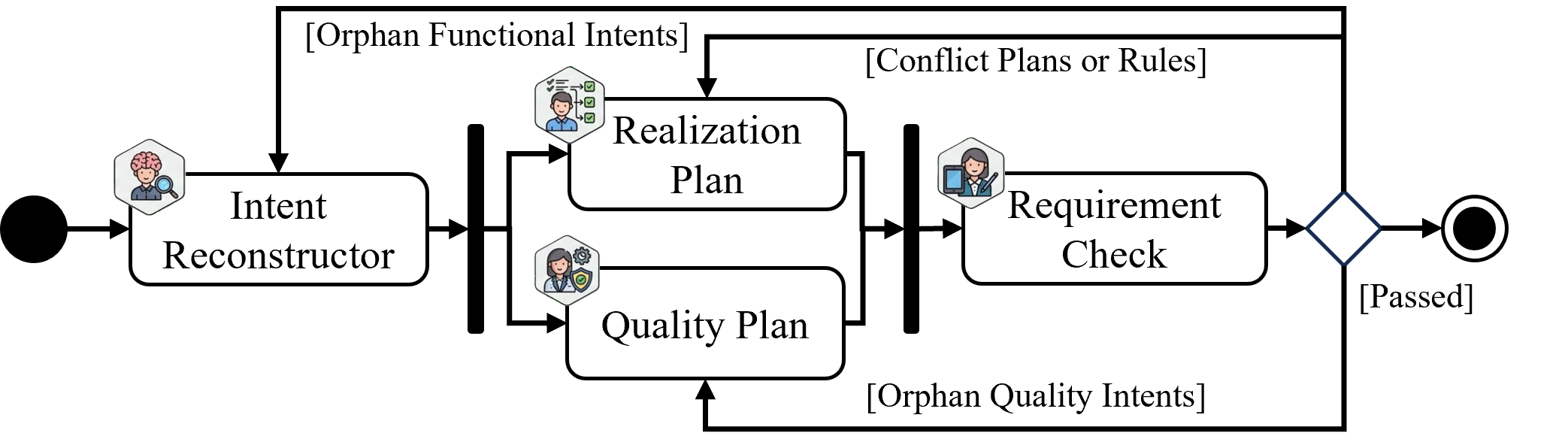}
    \caption{Activity Diagram of Bidirectional Process}
    \label{fig:collaboration_mechanism}
\end{figure}

\textbf{Multi-rounds Iterative Optimization:} Rather than a linear pipeline, the system utilizes a closed-loop, iterative feedback mechanism as illustrated in Figure \ref{fig:collaboration_mechanism}. Following the initial construction of the \textbf{BRT} by \textit{Intent Reconstructor}, \textit{Quality Planner} and \textit{Realization Planner} execute concurrently to develop the functional and quality parts of the \textbf{BRT}. Then, \textit{Requirement Checker} evaluates the \textbf{BRT} and dispatches localized feedback: Orphan Functional Intents are routed back to the \textit{Intent Reconstructor}, Orphan Quality Intents to \textit{Quality Planner}, and Conflict Plans/Rules to \textit{Realization Planner} for targeted resolution. This mechanism equips the framework with dynamic self-correction capabilities, guaranteeing the structural reliability and safety of the generated result.

\textbf{Convergence and Termination Criteria.} \textit{Requirement Checker} orchestrates the convergence of this iterative process. Successful termination strictly requires the \textbf{BRT} to exhibit zero orphan nodes and no logical conflicts. To prevent infinite loops, we enforce a maximum threshold of three iterations. Pilot experiments reveal that exactly three rounds provide an optimal balance between computational overhead and completion performance. Ultimately, these criteria ensure the system reliably outputs complete and safe rules.

\subsection{Definition of BRT}
The \textbf{BRT} serves as the core of our method. It explicitly captures the hierarchical relationships among user intents, logical plans, and concrete TAP rules.

\begin{figure*}
    \centering
    \includegraphics[width=1\linewidth]{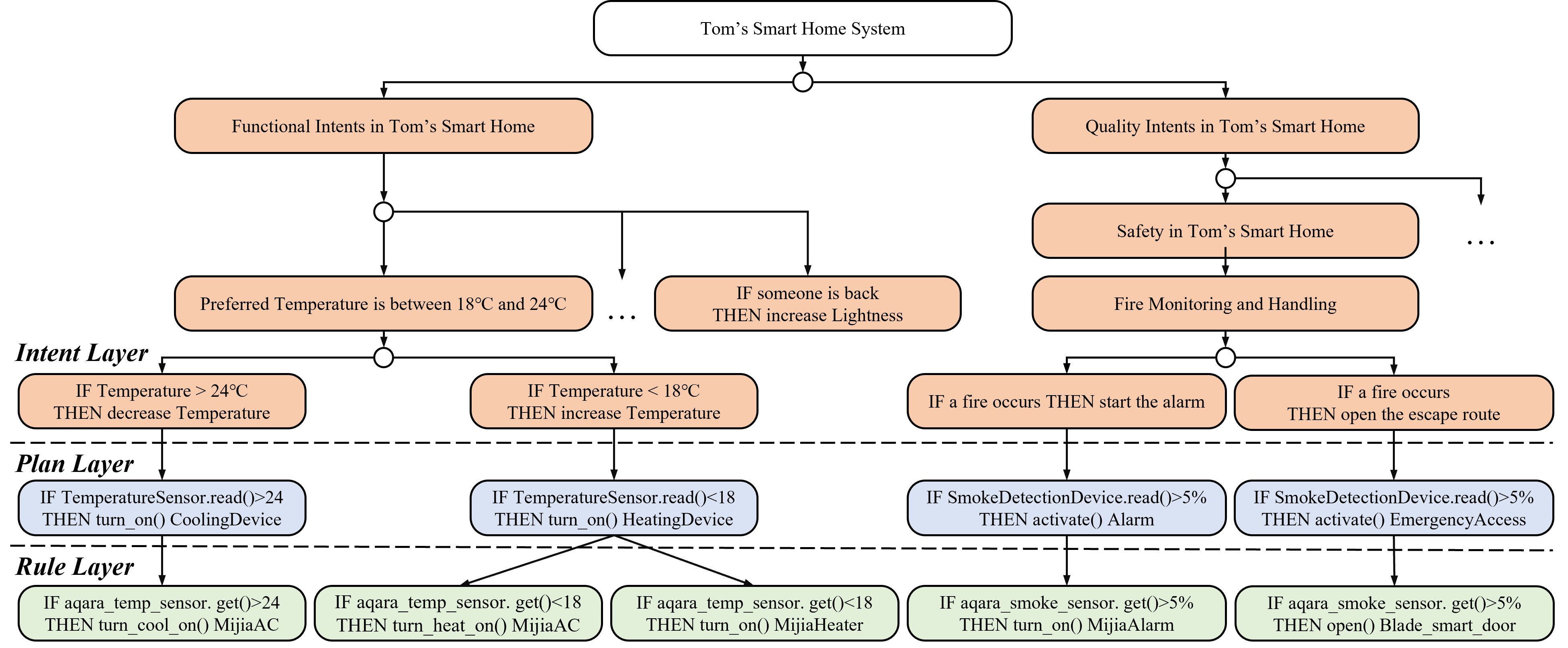}
    \caption{A Specific BRT Example of TOM's Smart Home System}
    \label{fig:BRT_Example}
\end{figure*}

\subsubsection{Formal definition}
\textbf{BRT} is formally defined as a three-layer directed tree structure, denoted as $T=(V,E)$, where $V$ represents the set of nodes and $E$ represents the set of edges. {Nodes} ($V = V_I \cup V_P \cup V_R$) comprise three distinct types:
\begin{itemize}
    \item Intent Layer Nodes ($V_I$): represents the user's desired intentions at the top layer. Each intent node encapsulates a device-agnostic goal, such as the functional intent node ``Preferred Temperature is between 18°C and 24°C'' and the safety intent node ``Fire Monitoring and Handling'', as illustrated in Figure \ref{fig:BRT_Example}.
    \item Plan Layer Nodes ($V_P$): contains logical planning schemes for a given intent. High-level intents are translated into causal, abstract device expressions (e.g., ``IF TemperatureSensor.read() > 24 THEN turn\_on() CoolingDevice''). Plan nodes act as a vital bridge, decomposing ambiguous high-level intentions into explicit causal expressions.
    \item Rule Layer Nodes ($V_R$): form the bottom layer of the \textbf{BRT} and represent concrete TAP rules, such as ``IF aqara\_temp\_-sensor.get() > 24 THEN turn\_cool\_on() MijiaAC''. Rule nodes serve as executable specifications that can be directly deployed to IoT devices.
\end{itemize}

\textbf{Edges} ($E \subseteq V \times V \times T$) define hierarchical decomposition relationships of type $t \in \{AND, OR\}$.
Formally, an edge $e = (v_{parent}, v_{child}, t) \in E$ indicates that $v_{child}$ is a decomposed sub-node of $v_{parent}$ under type $t$.
To preserve structural integrity, valid connections are constrained to a hierarchy: $(v_{parent}, v_{child}) \in (V_I \times V_I) \cup (V_I \times V_P) \cup (V_P \times V_R)$. This enforces a systematic translation pipeline that refines high-level intents into granular sub-intents, maps leaf intents to abstract logical plans, and ultimately grounds plans into concrete, device-specific rules.
The logical satisfiability of a parent node is strictly governed by $t$. An AND-decomposition ($t = AND$) requires joint realization; $v_{parent}$ is satisfied if and only if $\forall(v_{parent}, v_{child}, AND) \in E$, $v_{child}$ is satisfied (e.g., ``Preferred Temperature is between 18°C and 24°C'' requires simultaneously satisfying both upper and lower boundary intents as shown in Figure \ref{fig:BRT_Example}). Conversely, an OR-decomposition ($t = OR$) provides alternative implementation pathways; $v_{parent}$ is satisfied if and only if $\exists(v_{parent}, v_{child}, OR) \in E$, $v_{child}$ is satisfied (e.g., an abstract heating plan is fulfilled by successfully activating either a smart air conditioner or a standalone heater).

In particular,\textbf{ BRT} is able to explicitly embed quality concerns as integral components of the intent layer by introducing a distinct subclass of quality intents ($V_Q \subset V_I$). By formalizing these quality concerns early, the \textbf{BRT} ensures that quality intents are no longer treated as passive, post-hoc verification checks; instead, they are elevated to core elements that inherently permeate the entire completion process.

\vspace{-0.1cm}
\subsubsection{Bidirectional Traceability Semantics}
The core value of BRT lies in its inherent bidirectional traceability, which aligns with the fundamental nature of intent-driven rule completion. Facilitated by hierarchical decomposition, the \textbf{BRT} enables semantic mapping between user intentions and concrete TAP rules through the bidirectional traceability semantics.

\textbf{Upward Traceability.} Given fragmented user rules, $R_{frag} \subseteq V_R$, upward traceability constructs a reverse ``rule $\rightarrow$ plan $\rightarrow$ intent'' derivation path to uncover the underlying objectives, effectively answering the ``why'' behind partial user inputs.
Driven by \textit{Intent Reconstructor}, the system traverses bottom-up to infer missing ancestral nodes. For example, a TAP rule like ``IF aqara\_temp\_-sen
sor.get() > 24 THEN turn\_cool\_on() MijiaAC'' is first abstracted into a logical plan (``IF TemperatureSensor.read() > 24 THEN turn\_-on() CoolingDevice''). This reveals an atomic intent to decrease the temperature when temperature is greater than 24℃. Leveraging common knowledge that environment temperature control necessitates dual bounds, the system deduces the overarching composite intent: ``Preferred Temperature between 18°C and 24°C'' This reverse chain formally reconstructs the true objectives concealed behind the user's fragmented rules.

\textbf{Downward Traceability.} Conversely, downward traceability addresses ``how'' to robustly and completely realize these intents. Initiating from the reconstructed intent nodes ($V_I$), Realization and \textit{Quality Planner} traverse top-down along the ``intent $\rightarrow$ plan $\rightarrow$ rule'' trajectory to systematically unfold comprehensive implementation schemes. Building upon the inferred ``Preferred Temperature between 18°C and 24°C'' intent, the system identifies the complementary lower-bound requirement (``IF Temperature < 18°C THEN increase Temperature''). This atomic intent is translated into an logical plan, which is subsequently OR-decomposed into alternative, executable rules—such as triggering the MijiaAC's heating mode or activating a standalone MijiaHeater as shown in  Figure \ref{fig:BRT_Example}.

Bidirectional traceability has a closed-loop characteristic: the intention nodes reconstructed via bottom-up traceability serve as the starting points for top-down generation, while the generated rule sets can be continuously verified through bottom-up traceability to ensure they faithfully reflect the original intentions. This closed-loop characteristic provides a rigorous guarantee for both the explainability and correctness of the rule completion process.

\section{Agents Implementation}
This section details the implementation of the four agents.

\vspace{-0.3cm}
\subsection{Technology Selection and Prompt Design}

\subsubsection{Technology Selection}
To fulfill the inference and generation requirements of the dual intent-driven requirement completion process, we utilize LLM as the core reasoning engines of agents. The selected foundation model must satisfy four critical criteria: (1) Generalization across Tasks: Seamlessly transitioning between bottom-up intent inference and top-down rule generation; (2) Semantic Parsing: Accurately understanding and extracting TAP rules, including devices, operations, and conditional thresholds; (3) Structured Output: Reliably producing JSON-formatted results that strict-ly conform to the \textbf{BRT}; and (4) Few-Shot Learning: Adapting to complex tasks via prompt without requiring extensive fine-tuning.

Besides, we construct a robust technical stack around agents. To bridge the domain knowledge gap inherent in general-purpose LLMs, we implement an Environment-Aware Augmented Generation mechanism. Rather than hardcoding static knowledge, the system dynamically retrieves active device profiles from current environment and contextually injects them into the prompts. This approach explicitly grounds the LLM's reasoning, mitigating domain-specific hallucinations while preventing context window bloat.

Finally, a downstream Output Validation module is introduced to ensure structural integrity of LLM's output. It conducts entity feasibility checks—verifying device existence, capabilities, and reasonable parameter values. Furthermore, it validates the structural mountability of generated components against the \textbf{BRT} to prevent orphan nodes or unintended overwrites. Validated outputs are standardized into correct structure, ensuring the output is syntactically compliant, structurally sound, and safely executable.

\vspace{-0.1cm}
\subsubsection{Prompt Design Framework}

To address the specific scenario of the TAP rule completion, we’ve devised a unified prompt template. The prompts for all four intelligent agents adhere to this structured framework, differing only in the specific content populated within the template. The template includes:

\textbf{Role Definition \& Scenario Description:} defines each agent's specific role and provides scenario description, helping the agent understand the particularity and background of IoT scenarios. It ensures that agent has a comprehensive grasp of the domain.

\textbf{Context Injection:} dynamically injects knowledge of current environment—such as device profiles—to ground the LLM's reasoning in the physical environment. In  particular, the real-time state of \textbf{BRT}, including existing intent, plan, and rule nodes, is embed into prompt. This structural context ensures new outputs remain consistent with the hierarchy, preventing structural conflicts.

\textbf{Reasoning \& Feedback:} mandates Chain-of-Thought (CoT\cite{wei2022chain}) and few-shot demonstrations for step-by-step reasoning. It also reserves an interface to inject inspection result from \textit{Requirement Checker} to iteratively correct problems.

\textbf{Constraints \& Boundaries:} hardcodes strict operational limits to ensure viability, such as prohibiting hallucinated device capabilities.
These predefined guardrails act as a safety layer.

\textbf{Reflection \& Structured Output:} guides the LLM through a self-reflection process to verify its logic against the constraints. Finally, it mandates a strict JSON output tailored to the specific agent's role, enabling seamless parsing.

\begin{figure}
    \centering
    \includegraphics[width=1\linewidth]{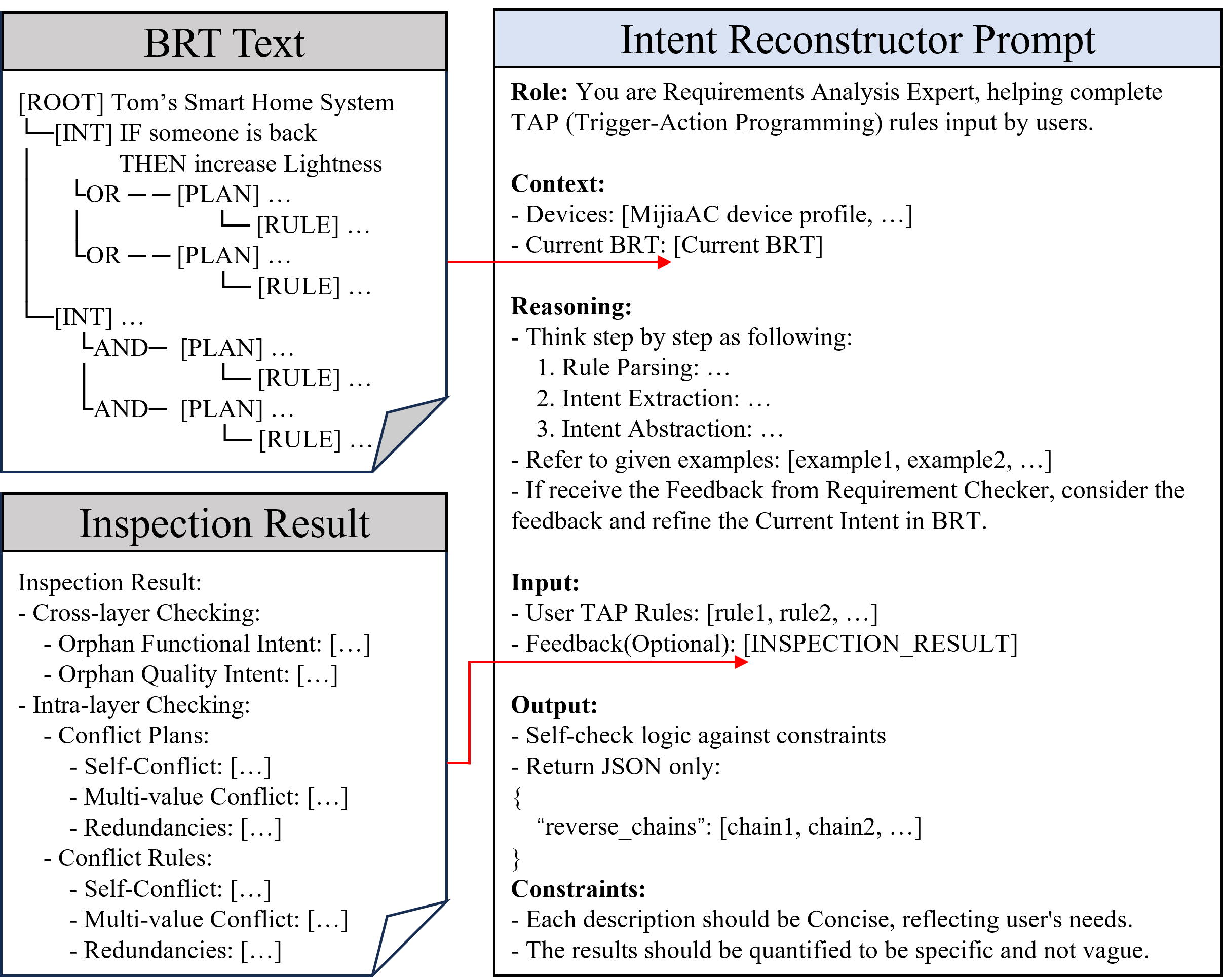}
    \caption{The Example of Intent Reconstructor's Prompt}
    \label{fig:promptInstance}
    \vspace{-0.3cm}
\end{figure}

\subsection{Intent Reconstructor Agent}

\textit{Intent Reconstructor Agent} is responsible for reverse-engineering incomplete TAP rules input by users to infer their underlying true functional intentions. The agent reconstructs the reverse three-layer requirement chain (rule → plan → intent) and populates the corresponding nodes and decomposition edges into the \textbf{BRT}.  The agent takes as input the user's fragmented TAP rules, the current state of the traceability tree, and optional feedback(containing orphan functional intents) from \textit{Requirement Checker}. The agent outputs planning nodes, intention nodes, and decomposition edges. The temperature parameter is set to 0.3.

\textbf{Task Decomposition.} 
The main process is decomposed into three steps. 
Step 1 (Rule Parsing) maps hardware-specific rules to platform-agnostic logical plans (e.g., translating a raw rule that turns cool on a specific MijiaAC triggered by aqara\_temp\_sensor into a generic plan to activate CoolingDevice by abstract TemperatureSensor). 
Step 2 (Intent Extraction) translates these plans into atomic intentions based on directly physical effects; turn on the CoolingDevice will ``Decrease Temperature'' for example. 
Step 3 (Intent Abstraction) employs a self-reflection mechanism to abstract atomic intentions into high-level composite intents according to common sense, ultimately identifying the user's overarching goal like ``Preferred Temperature between 18℃ and 24℃''.

\textbf{Prompt Design.} As show in Figure \ref{fig:promptInstance}  The prompt sets the agent as a Requirements Analysis Expert and provides some background knowledge of IoT scenerios, such as the full name of TAP. It injects the current \textbf{BRT} state dumped as text and domain knowledge such as ``MijiaAC device profile'' in Device Memory. The three-step decomposition is embedded as explicit instructions in Reasoning part. To facilitate in-context learning, the prompt includes some examples demonstrating how to extrapolate high-level intents from fragmented rules. A self-reflection mechanism requires the agent to check its reasoning process before the final result with the output format constrained to tailored JSON schema. For iterative correction, feedback from \textit{Requirement Checker}is injected to guide revision of only the problematic parts in current \textbf{BRT}.

\subsection{Quality Planner Agent}

\textit{Quality Planner Agent} acts as a dynamic injector of quality intentions, supplementing the \textbf{BRT} with essential quality requirements derived from quality template and environmental constraints. It constructs a forward three-layer chain (quality intent → plan → rule) to enhance \textbf{BRT}. The agent input general quality template and devices profile, and optional inspection result from \textit{Requirement Checker}. The agent outputs quality intention nodes and corresponding implementation paths. To ensure high precision and minimize hallucinations, the temperature is set to 0.2.

\textbf{Task Decomposition.} The main process is decomposed into two steps. Step 1 (Template Instantiation) intersects expert-author-ed quality templates with the current device profiles and the realizable part will be discarded subsequently. For instance, a general safety template might state: ``If a fire occurs THEN start the alarm''. The agent evaluates environmental feasibility to successfully instantiate this into a logical plan; if a smoke sensor and an alarm are available, the corresponded plan will be kept. When exact hardware matches are absent, the agent will start heuristic reasoning from the general template to infer expected quality intentions and logical plans.
Step 2 (Rule Generation) maps logical plans to specific devices to generate executable TAP rules, such as the previous plan ``IF SmokeDetectionDevice.read()>5\% THEN activate() Alarm'' to concrete TAP rule ``IF aqara\_smoke\_sensor.get() > 5\% THEN turn\_on() MijiaAlarm''.

\textbf{Prompt Design.} The prompt defines the agent as an IoT Safety and Optimization Expert. It adheres to our unified framework by injecting the current \textbf{BRT} State and Device Profiles as context. Embedding step-by-step task decomposition, the agent is forced to carefully consider each quality intentions before generating a structured JSON output. For iterative correction, orphan quality intents is fed back to guide the agent in making revisions. To prevent the generation of unsupported devices, the prompt strictly outlines the physical boundaries of the current smart environment via hardcoded constraints. Furthermore, a self-reflection mechanism requires the agent to
verify the chain and constraints satisfiability after generation to mitigate hallucinations.

\subsection{Realization Planner Agent}

\textit{Realization Realization Planner} is responsible for intention realization, translating abstract functional intentions within the \textbf{BRT} into concrete device scheduling logic progressively.
The agent realizes the forward three-layer requirement chain (quality intent → plan → rule). New intents, optional feedback from \textit{Requirement Checker} are input. The agent finally outputs the implementation paths of targeted intentions. The temperature parameter is set to 0.3.

\textbf{Task Decomposition.} The main process is decomposed into three steps. Step 1 (Intent Decomposition) refines high-level intents into actionable atomic intents. Step 2 (Plan Design) formulates multiple candidate logical plans for each atomic intent. Step 3 (Rule Generation) grounds these plans into executable TAP rules bound to specific device capabilities as \textit{Quality Planner}. 

\textbf{Prompt Design.} Defined as a IoT Implementation Expert, the agent utilizes our unified prompt framework to dynamically inject real-time \textbf{BRT} states and device profiles. It employs task decomposition to logically navigate the complex generation process. The agent strictly outputs a JSON-formatted three-layer chain comprising intent, plan, and rule nodes. When receiving plan or rule conflicts, a dynamically refinement will be activated until zero errors remain or a predefined maximum iteration threshold is reached. A self-reflection similar to \textit{Quality Planner} is infused as well.

\subsection{Requirement Checker Agent}

\textit{Requirement Checker Agent} acts as the method's final quality gatekeeper, conducting comprehensive structural and semantic inspections of the \textbf{BRT}. Because multi-agent collaboration can inadvertently introduce intra-layer logical contradictions or inter-layer fulfillment gaps, this agent is essential for guaranteeing the completeness, consistency, and safe deployability of the final TAP rules. The agent input current \textbf{BRT} and output final inspection result.

\textbf{Task Decomposition.} The main process is decomposed into two parallel steps. Step 1 (Cross-Layer Checking) identifies unfulfilled intents by tracing forward to ensure high-level intentions are fully covered by lower-layer plans, while tracing backward to verify that bottom-tier implementations strictly align with the overarching intents. Step 2 (Intra-Layer Checking) detects logical conflicts among plans and rules using a two-stage hybrid approach. To resolve ``formally valid but semantically absurd'' errors, the agent first leverages common knowledge to map device capabilities into normalized semantic categories and subsequently applies formal verification (e.g., TAPChecker~\cite{chen2024tapchecker}) to strictly identify self-conflicts (rules that can never execute), multi-value conflicts (contradictory commands sent to the same device), and redundancies.

\textbf{Prompt Design.} Acting as a rigorous Requirement Auditor, the agent is prompted to evaluate the normalized logic against the unified \textbf{BRT} context. It outputs a structured, JSON-formatted diagnostic report explicitly identifying unfulfilled user intents, conflicting plans and rules. These reports are persistently stored in the workspace to guide other agents' refinement iterations.


\section{Experimental Evaluation}
To evaluate the effectiveness of our method, we conducted a large-scale study to answer three research questions.

\label{section4-1}
\vspace{-0.1cm}
\subsection{Research Questions}

\textbf{RQ1 (Performance Comparison):} How does our method's completion performance compare with other methods in end-user requirements completion? This question aims to evaluate the effectiveness of our method in end-user's requirement completion tasks.

\textbf{RQ2 (Ablation Study):} What are the individual performance contributions of the distinct agents and mechanisms? This question investigates the specific contributions of different components.

\textbf{RQ3 (Cost \& Efficiency):} How do the computational latency and financial costs of our method compare to existing baselines? This question examines the method's viability for IoT deployment.

\subsection{Datasets}
\label{section4-2}

The datasets selected for our experiments satisfy two key criteria: 
(i) \textbf{Diversity in complexity and scenarios}: The datasets exhibit a range of task complexities, from simple to highly complex user intentions and device management. Besides, the datasets cover six common smart home scenarios including kitchen, study room, living room, bedroom, dining room, and bathroom.
(ii) \textbf{Representativeness of real-world requirements}: In particular, the constructed datasets are sourced from genuine real-world data. Our datasets are sourced from real-world data from IFTTT. The data were independently annotated by three domain experts and cross-validated to ensure annotation quality and reliability.


The datasets are formed by five levels, 292 cases, 2938 TAP rules in total. The details of our benchmark is shown in Table \ref{tab:benchmark_info}. The inclusion of these datasets ensures that our evaluation encompasses not only a broad spectrum of scenarios and complexity levels but also the practical challenges encountered in smart home tasks.

\begin{table}[!t]
  \caption{Statistics of Our Benchmark. ``Multi-Impl.'' stands for cases with multiple implementation options}
  \label{tab:benchmark_info}
  \vspace{-0.15in}
  \centering
  \setlength{\tabcolsep}{4pt} 
  \begin{tabular}{cccccc}
    \toprule
    Dataset & \#Case & \makecell{\#Multi-\\Impl.} & \makecell{\#Rule} & \#Device & Commonality \\         
    \midrule
    L1 & 56 & 0  & 1-12  & 1-13  & Common \\
    L2 & 66 & 5  & 1-14  & 2-13  & Common \\
    L3 & 57 & 11 & 2-21  & 3-14  & Most Common \\
    L4 & 56 & 17 & 10-24 & 9-17  & Partial Common \\
    L5 & 57 & 20 & 11-33 & 10-18 & Most Uncommon \\
    \bottomrule
  \end{tabular}
  \vspace{-0.20in}
\end{table}

\subsection{Evaluation Metrics}
\label{section4-3}

To evaluate the performance of our approach, we adopt two categories of metrics: intent-level metrics, which assess the correctness and diversity of implemented user intents, and rule-level metrics, which evaluate the completeness, hallucination, and correctness of the generated TAP rules.

For intent-level evaluation, we employ two metrics, that is, Intent Implementation Correctness Rate (IIC) and Intent Implementation Diversity Rate (IID). Among them, \textbf{Intent Implementation Correctness Rate(IIC)} measures the proportion of user intents correctly implemented by the generated TAP rules compared to the ground truth.
For a given intent that may be realized through multiple implementation in the ground truth, it is considered covered if at least one of its corresponding implementation groups is matched by the generated rules.
\begin{equation}
    IIC = \frac{\#\text{Correctly Implemented Intents}}{\#\text{User Intents}}
\end{equation}

\textbf{Intent Implementation Diversity Rate (IID)} measures, among those intents that can be realized through multiple implementation paths, the proportion of implementations that are fully implemented. It reflects the fault tolerance of intent realization and serves as an indicator of system vulnerability.

\begin{equation}
    IID = \frac{\#\text{Fully Implemented Intents}}{\#\text{Multi-Implementation Intents}}
\end{equation}

For rule-level evaluation, we introduce three metrics: Rule Completion Rate (RCmpR), Rule Hallucination Rate (RHR), and Rule Conflict Rate (RConR). \textbf{Rule Completion Rate (RCmpR)} evaluates the completeness of the generated TAP rules, excluding hallucinated ones. It reflects how many valid TAP rules are successfully completed compared to the ground truth. 

\begin{equation}
    RCmpR = \frac{\#\text{Successfully Completed TAP Rules}}{\#\text{TAP Rules Needing to be Completed}}
\end{equation}

\textbf{Rule Hallucination Rate (RHR)} quantifies the proportion of generated TAP rules that are irrelevant. 
\begin{equation}
    RHR = \frac{\#\text{Irrelevant TAP Rules}}{\#\text{Generated TAP Rules}}
\end{equation}

\textbf{Rule Conflict
Rate (RConR)} quantifies the proportion of generated TAP rules that are incorrect. 
\begin{equation}
    RConR = \frac{\#\text{Incorrect TAP Rules} }{\#\text{Generated TAP Rules}}
\end{equation}

Finally, we define the \textbf{Correctness Criteria} for TAP rules. A rule is considered correct if it satisfies syntactic validity and entity feasibility. Specifically, generated rules must conform to the TAP rule syntax and correctly reference the provided devices and their functions. Rules beyond those present in the ground truth are considered valid as long as they meet the syntactic and entity criteria.
\vspace{-0.2cm}
\subsection{Baselines}
\label{section4-4}
To rigorously evaluate the effectiveness of our approach across different system architectures, we selected six representative baselines as shown in Table ~\ref{tab:baselines}. This selection was guided by four core principles: (i) \textbf{comprehensive technical coverage:} spanning deep learning, LLMs, and agent-based methods; (ii) \textbf{domain specificity:} balancing IoT-specific and general-purpose systems; (iii) \textbf{accessibility:} preferring open-source methods, and (iv) \textbf{timeliness}. Specifically, our baselines include IoT-tailored models like the LLM-based ChatIoT~\cite{gao2024chatiot} and the deep learning approach TARGE~\cite{cimino2025iot}, alongside a general-purpose LLM reasoning paradigm LLM+CoT~\cite{wei2022chain} to isolate domain-specific design influences. To assess the performance of agentic architectures in IoT scenarios, we also incorporate state-of-the-art single-agent systems—Manus~\cite{Manus} and its open-source counterpart, openManus~\cite{OpenManus}—as well as the role-based multi-agent collaboration framework, MetaGPT~\cite{hong2023metagpt}. This diverse selection ensures a comprehensive comparison of our multi-agent framework against both specialized and general-purpose alternatives.

\begin{table}[h]
\centering
\caption{The Overall information of Different Baselines}
\label{tab:baselines}
   \vspace{-0.15in}
\resizebox{\columnwidth}{!}{
  \begin{tabular}{ccccc}
    \toprule
    \textbf{Method Name} & \textbf{Type} & \textbf{Is IoT-Specfic} & \textbf{Open Source} \\
    \midrule
    ChatIoT & LLM-based Framework & Yes & Open \\
    TARGE & Deep Learning & Yes & Open  \\
    LLM+CoT  & Prompt Engineering & No & Closed  \\
    openManus & General-Purpose Agent & No & Open  \\
    Manus & General-Purpose Agent & No & Closed  \\
    MetaGPT & Multi-Agent & No & Open \\
    \bottomrule
  \end{tabular}%
}
\end{table}
\vspace{-0.2cm}

\vspace{-0.2cm}
\section{Results and Analysis}
\label{section5}

\subsection{RQ1: Performance Comparison}
\label{section5-1}
\begin{table*}[!t]
\centering
\caption{Performance Comparison of Methods}
\label{tab:RQ1_result_table}
\vspace{-0.15in}
\renewcommand{\arraystretch}{0.9}
\begin{tabularx}{\textwidth}{
@{}l *{9}{>{\centering\arraybackslash}X}@{}
}
\toprule
DataSet & Metrics & ChatIoT(\%) & TARGE(\%) & LLM+CoT(\%) & OpenManus(\%) & Manus(\%) & MetaGPT(\%) & Ours(\%) & Gain(\%)\\
\midrule
\multirow{5}{*}{L1}
 & IIC & 36.91 & 39.88 & 54.01 & 59.63 & 63.78 & 54.94 & \textbf{79.01} & \textbf{$\uparrow$15.23} \\
 & IID & - & - & - & - & - & - & - & -\\
 & RCmpR & 49.49 & 53.01 & 58.69 & 60.74 & 67.56 & 57.53 & \textbf{82.89} & \textbf{$\uparrow$15.33}\\
 & RHR & 0 & 66.27 & 36.76 & 57.82 & 42.56 & 34.74 & \textbf{12.51} & \textbf{$\downarrow$22.23}\\
 & RConR & 0 & 43.04 & 19.09 & 28.57 & 20.24 & 12.42 & \textbf{0} & \textbf{$\downarrow$12.42}\\
\midrule
\multirow{5}{*}{L2}
 & IIC & 23.36 & 27.79 & 39.40 & 44.41 & 44.91 & 39.19 & \textbf{75.10} & \textbf{$\uparrow$30.19}\\
 & IID & 0 & 0 & 17.50 & 35.00 & 37.50 & 17.50 & \textbf{65.00} & \textbf{$\uparrow$27.5}\\
 & RCmpR & 22.71 & 29.06 & 36.36 & 41.38 & 42.33 & 39.50 & \textbf{78.38} & \textbf{$\uparrow$36.05}\\
 & RHR & 0 & 81.27 & 57.80 & 69.14 & 62.92 & 49.50 & \textbf{25.13} & \textbf{$\downarrow$24.37}\\
 & RConR & 0 & 50.62 & 22.54 & 40.67 & 24.91 & 14.53 & \textbf{0} & \textbf{$\downarrow$14.53}\\
\midrule
\multirow{5}{*}{L3}
 & IIC & 18.80 & 24.50 & 32.27 & 33.07 & 33.15 & 31.52 & \textbf{71.55} & \textbf{$\uparrow$38.40}\\
 & IID & 0 & 0 & 9.09 & 0 & 11.76 & 11.36 & \textbf{61.36} & \textbf{$\uparrow$49.60}\\
 & RCmpR & 15.03 & 22.14 & 28.05 & 32.07 & 32.56 & 31.52 & \textbf{73.49} & \textbf{$\uparrow$40.93}\\
 & RHR & 0 & 91.09 & 61.74 & 69.84 & 65.81 & 50.35 & \textbf{27.64} & \textbf{$\downarrow$22.71}\\
 & RConR & 0 & 77.65 & 25.42 & 45.44 & 30.31 & 22.38 & \textbf{4.48} & \textbf{$\downarrow$17.90}\\
\midrule
\multirow{5}{*}{L4}
 & IIC & 18.26 & 19.45 & 29.10 & 31.52 & 31.57 & 25.31 & \textbf{62.46} & \textbf{$\uparrow$30.89}\\
 & IID & 0 & 0 & 0 & 0 & 0 & 0 & \textbf{53.37} & \textbf{$\uparrow$53.37}\\
 & RCmpR & 10.58 & 11.21 & 21.47 & 26.68 & 26.82 & 23.69 & \textbf{68.84} & \textbf{$\uparrow$42.02}\\
 & RHR & 0 & 93.63 & 73.98 & 78.10 & 71.96 & 54.09 & \textbf{39.56} & \textbf{$\downarrow$14.53}\\
 & RConR & 0 & 79.30 & 26.33 & 46.82 & 32.52 & 24.70 & \textbf{6.35} & \textbf{$\downarrow$18.35}\\
\midrule
\multirow{5}{*}{L5}
 & IIC & 17.98 & 19.41 & 27.25 & 30.61 & 31.42 & 24.44 & \textbf{63.15} & \textbf{$\uparrow$31.73}\\
 & IID & 0 & 0 & 0 & 0 & 0 & 0 & \textbf{38.36} & \textbf{$\uparrow$38.36}\\
 & RCmpR & 10.02 & 11.10 & 20.49 & 26.67 & 26.01 & 20.48 & \textbf{69.67} & \textbf{$\uparrow$43.00}\\
 & RHR & 0 & 93.77 & 73.82 & 79.41 & 72.83 & 58.94 & \textbf{40.41} & \textbf{$\downarrow$18.53}\\
 & RConR & 0 & 79.73 & 28.55 & 47.86 & 35.30 & 30.96 & \textbf{7.44} & \textbf{$\downarrow$21.11}\\
\bottomrule
\end{tabularx}
\end{table*}

\textbf{Setup.} To evaluate our method, we conducted comparative experiments against baselines on a self-constructed 5-level difficulty dataset under a unified assessment protocol. For LLM-based methods (ChatIoT, LLM+CoT, openManus, MetaGPT), we use Deepseek-V3.2~\cite{liu2025deepseek} as the foundation model to ensure fairness, with the Temperature parameter set to 0.3 and top\_p set to 1. Besides, TARGE was fine-tuned using its official pre-trained model. 

\textbf{Results.} As detailed in Table \ref{tab:RQ1_result_table}, our method consistently achieves state-of-the-art performance across all five difficulty levels and evaluation metrics. All baselines show performance cliffs as task complexity scales, contrasting with our method's robust stability. On the most challenging L5 tasks, our method still maintains an IIC of over 63\%, effectively about doubling the performance of the best baseline, Manus. Most strikingly, while all baseline methods completely fail to generate diversified implementation schemes in highly complex scenarios (scoring zero for IID at L4 and L5), our approach sustains a diversity rate exceeding 38\% in L5. Regarding rule generation, our method achieves a RCmpR near 70\%, outperforming the next best baseline by 43 percentage points. Furthermore, excluding ChatIoT which maintains zero hallucination and conflict rates by strictly modifying existing rules rather than generating new ones, resulting in an about 10\% completion rate, our method exhibits well safety. It significantly suppresses both RHR and RConR, dropping them below the best-performing baselines.

\textbf{Analyses.} 
A qualitative analysis reveals that our method's outstanding performance across all complexity levels stems from three core design points rather than a simple accumulation of prompts.

First, our method systematically integrates quality intents into the completing process. Rather than passively relying on statistical co-occurrence or generalized common sense, our \textbf{BRT} explicitly consider all intentions, capturing quality concerns alongside functional intentions. This proactive approach elevates the generated rules from basic functional mapping to holistic trustworthiness. For instance, in a smart TV scenario, our approach inherently prioritizes and generates critical safety rules, such as automatically turning off appliances when leaving home, which is often overlooked by other methods.

Second, as an added benefit, our method naturally supports implementation diversity. By utilizing explicit AND/OR decomposition within the BRT, our framework inherently models flexible execution paths (e.g., managing temperature via either a heater or an air conditioner). This structural awareness allows our method to consistently formulate multi-path solutions rather than relying on a single rigid path, directly yielding our non-zero IID scores.

Finally, our framework is specifically designed to neutralize hallucinations and conflicts through a triple-layered defense. In complex scenarios, some method are prone to function drift (recommending unrelated functions like ``air circulation'' because they co-occur on the same air conditioner) or excessive creation (generating unsolicited rules that deviate far from user’s input). Crucially, these hallucinated rules frequently clash with existing user input rules, resulting in contradictory commands (e.g., simultaneously attempting to brighten and turn\_on the exact same light) and causing RConR to spike. We mitigate these risks via upward intent tracing to maintain alignment with the user's intentions, environment-aware realization to keep device actions consistent with intentions, and formal conflict checking to eliminate contradictory commands. These mechanisms guarantee a safe, consistent, and intent-aligned rule set, which explains our drastically reduced RHR and RConR.

\vspace{-0.1cm}
\begin{center}
\fcolorbox{black}{gray!10}{\parbox{.95\linewidth}{\textbf{Answer to RQ1:} Our bidirectional traceability approach significantly outperforms baselines, improving completion effects by an absolute 43\% while reducing conflicts by an absolute 21\%.}}

\end{center}

\begin{table*}[t!]
\centering
\caption{Result of Ablation Study}
\label{tab:RQ2_result_table}
 \vspace{-0.15in}
\renewcommand{\arraystretch}{1}
\begin{tabularx}{0.98\textwidth}{@{}l *{10}{>{\centering\arraybackslash}X}@{}}
\toprule
 & IIC(\%) & & IID(\%) & & RCmpR(\%) & & RHR(\%) & & RConR(\%) & \\
\midrule
\textbf{Ours} & 71.55 & & 61.36 & & 73.49 & & 27.64 & & 4.48 &\\
\midrule

\textbf{w/o Agent} & & & & & & & & & & \\
\text{Intent Reconstructor} & 60.99 & \textcolor{red}{$\downarrow$ 10.56\%} & 19.70 & \textcolor{red}{$\downarrow$ 41.66\%} & 64.87 & \textcolor{red}{$\downarrow$ 8.62\%} & 27.81 & \textcolor{red}{$\uparrow$ 0.17\%} & 25.49 & \textcolor{red}{$\uparrow$ 21.01\%} \\
\text{Realization Planner} & 17.69 & \textcolor{red}{$\downarrow$ 53.86\%} & 0 & \textcolor{red}{$\downarrow$ 61.36\%} & 14.99 & \textcolor{red}{$\downarrow$ 58.50\%} & 76.77 & \textcolor{red}{$\uparrow$ 49.13\%} & 12.17 & \textcolor{red}{$\uparrow$ 7.69\%} \\
\text{Quality Planner} & 20.34 & \textcolor{red}{$\downarrow$ 51.21\%} & 0 & \textcolor{red}{$\downarrow$ 61.36\%} & 19.10 & \textcolor{red}{$\downarrow$ 54.39\%} & 68.76 & \textcolor{red}{$\uparrow$ 41.12\%} & 5.47 & \textcolor{red}{$\uparrow$ 0.99\%} \\
\text{Requirement Checker} & 57.78 & \textcolor{red}{$\downarrow$ 13.77\%} & 39.36 & \textcolor{red}{$\downarrow$ 22.00\%} & 57.10 & \textcolor{red}{$\downarrow$ 16.39\%} & 42.78 & \textcolor{red}{$\uparrow$ 15.14\%} & 33.91 & \textcolor{red}{$\uparrow$ 29.43\%} \\
\midrule
\textbf{w/o Mechanism} & & & & & & & & & & \\
\text{BRT} & 59.42 & \textcolor{red}{$\downarrow$ 12.13\%} & 0 & \textcolor{red}{$\downarrow$ 61.36\%} & 58.18 & \textcolor{red}{$\downarrow$ 15.31\%} & 28.13 & \textcolor{red}{$\uparrow$ 0.49\%} & 35.16 & \textcolor{red}{$\uparrow$ 30.68\%} \\
\text{Multi-rounds Iteration} & 68.98 & \textcolor{red}{$\downarrow$ 2.57\%} & 41.67 & \textcolor{red}{$\downarrow$ 19.69\%} & 72.12 & \textcolor{red}{$\downarrow$ 1.37\%} & 33.70 & \textcolor{red}{$\uparrow$ 6.06\%} & 19.11 & \textcolor{red}{$\uparrow$ 14.63\%} \\
\bottomrule
\end{tabularx}
\end{table*}

\vspace{-0.2cm}
\subsection{RQ2: Ablation Study}
\label{section5-2}

\textbf{Setup.} 
To evaluate individual contributions, we test six ablated variants on the L3 dataset using DeepSeek-V3.2: omitting the Intent Reconstructor, Quality Planner, Realization Planner, Requirement Checker, BRT (replaced by unstructured text), and Iteration (restricted to one round). All standard metrics are maintained.

\textbf{Results.} As detailed in Table \ref{tab:RQ2_result_table}, our evaluation highlights a clear ablation result driven by three critical metrics to validate our design. On IIC and RCmpR, \textit{Realization Planner} and \textit{Quality Planner} are the most indispensable components. Ablating them causes disastrous losses in completion effect, plunging IIC by over 53\% and 51\%, and RCmpR by over 58\% and 54\%, respectively. For the safety metric RconR, the \textbf{BRT} representation is the most critical structural foundation, as its removal induces a massive about 30\% surge in logical conflicts. Following this, bypassing the \textit{Requirement Checker} triggers a secondary catastrophic failure for safety, spiking RConR by over 29\%. Other metrics further validate this functional divide; for instance, IID collapses to zero without the core planners or the \textbf{BRT}, while RHR predictably worsen across all ablations.

\textbf{Analyses.} Qualitative analysis shows how component absences disrupt specific functions within the completion process. 

Regarding completion effects, \textit{Realization Planner} and \textit{Quality Planner} serve as the most indispensable drivers. Removing \textit{Realization Planner} severs the top-down grounding of abstract intents into concrete rules. This forces the system into direct mapping rather than progressive optimization, triggering the sharpest drop by over 53\% in RCmpR subsequently affecting the implementation of intentions. 
Ablating \textit{Quality Planner} exposes a ``safety black hole'' by ignoring implicit quality concerns, such as automatically turning off a heater when leaving, reducing systematic quality assurance to mere blind guessing and severely degrading completion effectiveness. 
Furthermore, \textit{Intent Reconstructor} is vital for bottom-up understanding; its absence leaves subsequent planning without a target anchor, further damaging intent correctness. IID also collapses without these agents or the \textbf{BRT}, as agents lose the structural context to navigate alternative paths.

In addition, the system relies on strict structure and formal validation.Removing the \textbf{BRT} strips the agents' shared semantic space, degrading collaboration into a black-box text relay and causing a massive 30.68\% surge in RConR. Bypassing \textit{Requirement Checker} triggers a surge in RConR by disabling global formal inspections. Without this final line of defense, well-designed components frequently introduce contradictions, such as different agents simultaneously issuing on/off commands to the same device. Correspondingly, RHR predictably spikes across these ablations, most notably when the normative planning agents are removed. Finally, without the multi-round iteration providing a closed-loop cycle to progressively correct deviations, residual hallucinations and hidden conflicts persist.

\begin{center}
\fcolorbox{black}{gray!10}{\parbox{.95\linewidth}{\textbf{Answer to RQ2:} While every component is indispensable, \textit{Realization Planner} and \textit{Quality Planner} avert over 50\% drops in intent correctness and rule completion, whereas the \textbf{BRT} and \textit{Requirement Checker} act as critical guardrails that suppress almost 30\% surges in conflicts.}}
\end{center}

\vspace{-0.2cm}
\subsection{RQ3: Cost \& Efficiency}
\label{section5-3}

\textbf{Setup.} To assess economic feasibility, we compared the average per-task execution times and inference costs of our method against LLM/Agent-based baselines. Tests utilized DeepSeek-V3.2, calculating costs via official token pricing.

\textbf{Results.} As detailed in Table \ref{tab:RQ3_result_table}, our method records an average financial cost of \$0.08056 per task and an average execution duration of 885.384 seconds. Compared to the baseline methods, which range in average cost from \$0.00070 to \$0.02300 and in average duration from 1.720 seconds to 278.788 seconds, our approach incurs higher resource overhead with high performance.

\textbf{Analyses.} Our method operates on a paradigm of ``higher investment yielding high-quality performance'' that is direct consequence of its rigorous, multi-layered architectural design. As shown in Table \ref{tab:RQ3_result_table}, all open-source agent-based methods have higher resource overheads than single LLM because of more complex thinking and actions. However, considering that rule completion is typically a low-frequency configuration rather than a high-frequency real-time operation, this average cost of roughly \$0.08 per task remains highly acceptable for end-users in safety-critical IoT scenarios, especially when contrasted with the recurring subscription fees of commercial smart home platforms such as  Google Home Premium\cite{google_store}.

\begin{center}
\fcolorbox{black}{gray!10}{\parbox{.95\linewidth}{\textbf{Answer to RQ3:} Our method incurs an average financial cost of approximately \$0.08 per task, trading a acceptable resource overhead for safe and high-quality rule completion.}}
\end{center}

\begin{table}[t!]
\centering
\caption{Average Cost and Duration of Different Methods}
\label{tab:RQ3_result_table}
\vspace{-0.15in} 
\renewcommand{\arraystretch}{0.9} 
\begin{tabularx}{\columnwidth}{@{}l XX@{}} 
\toprule
Methods & \centering Avg. Cost (\$) & \centering\arraybackslash Avg. Duration (sec) \\
\midrule
ChatIoT   & \centering 0.00122 & \centering\arraybackslash 14.952 \\
LLM+CoT   & \centering 0.00070 & \centering\arraybackslash 11.546 \\
OpenManus & \centering 0.00546 & \centering\arraybackslash 278.788 \\
Manus     & \centering 0.02300 & \centering\arraybackslash 1.720 \\
MetaGPT   & \centering 0.01036 & \centering\arraybackslash 16.704 \\
\textbf{Ours} & \centering \textbf{0.08056} & \centering\arraybackslash \textbf{885.384} \\
\bottomrule
\end{tabularx}
\vspace{-0.5cm}
\end{table}

\vspace{-0.1cm}
\section{Threats to Validity}
\label{section6}
\textbf{Internal Validity.} A primary threat involves the inherent non-determinism of the foundation LLM (DeepSeek-V3.2). To mitigate reasoning randomness, we enforced a low temperature setting acro-ss all LLM-based approaches. 
Furthermore, the quality templates utilized by \textit{Quality Planner} are authored by human experts; this inherent subjectivity may introduce bias and slightly limit the framework's zero-shot adaptability to unmodeled scenarios.
Additionally, despite employing a unified prompt framework, LLMs' inherent sensitivity to subtle semantic variations may introduce minor fluctuations in reasoning trajectories and rule generation.

\textbf{External Validity.} Our findings are grounded in a self-construct-ed benchmark. While it leverages real-world IFTTT data spanning six scenarios and five difficulty levels, it may not entirely encapsulate diverse global user habits. 
Besides, our method assumes well-defined device profiles, risking performance degradation in real-world IoT systems with legacy devices, closed APIs, or poor documentation.
Additionally, the framework's generalizability to broad-er programming paradigms, such as workflow, and its transferability to complex IoT system require further empirical validation.

\textbf{Construct Validity.} Defining and evaluating ``user intent'' involves inherent ambiguity and subjectivity. To ensure intent-level metrics accurately captured requirement fulfillment, we established ground truth through independent annotation and cross-validation by three experts, trying to minimize individual interpretive bias. 
Additionally, while our static evaluation metrics confirm rule correctness, they inherently overlook dynamic runtime behaviors(e.g., synchronization delays, network latency, and race conditions). Future in situ studies is necessary to validate practical safety.


\textbf{Multi-Agent System Validity.} A core challenge in our multi-agent framework is the risk of cascading error propagation and non-convergence. Each agent's hallucination may be amplified during collaboration. We mitigated this by using the \textbf{BRT} as a shared semantic anchor. We also implemented structured feedback for \textit{Requirement Checker} and enforced maximum iteration thresholds; empirically, most samples converged stably within three rounds.

\vspace{-0.3cm}
\section{Related Work}
\label{section7}

\textbf{Traditional and Deep Learning Approaches} treat TAP rule completion as a recommendation problem based on historical data; for example, Dominguez \emph{et al.}~\cite{dominguez2018mashup} leveraged the SVD++ mashup model, Corno \emph{et al.}~\cite{corno2019recrules} proposed RecRules using collaborative semantic graphs, and Liu \emph{et al.}~\cite{liu2022device} combined K-means clustering with the Apriori algorithm. To improve semantic understanding, subsequent deep learning models mapped natural language requirements into latent spaces, as demonstrated by the TARGE model by Cimino \emph{et al.}~\cite{cimino2025iot} and the MvTAP framework by Wu \emph{et al.}~\cite{wu2025user}. However, these approaches are constrained by their reliance on pattern-matching and historical data density, often failing to capture the deep semantics required for complex, open-domain logical reasoning. 

\textbf{LLM-Based Approaches} leverage strong contextual understanding and generation capabilities for direct TAP rule generation. Gao \emph{et al.}~\cite{gao2024chatiot} introduce ChatIoT, which employs a dialogue-based interaction framework with context-aware compressed prompting to process user requirements. Wu \emph{et al.}~\cite{wu2025llm4tap} propose LLM4TAP, enhancing user-rule interaction graphs via singular value decomposition and contrastive learning to extract deep user intents from rule texts. While these LLM-based methods offer impressive generative power, they introduce inherent hallucination issues. Models may fill in missing logical conditions based on training data, producing seemingly plausible but potentially unsafe rules, which poses serious physical safety risks in IoT deployment scenarios.

\textbf{Multi-Agent Approaches} mitigate the uncontrollable hallucinations of LLM by incorporating dynamic closed-loop feedback and role-based collaboration. Hong \emph{et al.}~\cite{hong2023metagpt} introduce Meta-GPT, leveraging human-standardized workflows to decouple complex tasks. Shen \emph{et al.}~\cite{shen2025mind} break down Manus~\cite{Manus}, which integrates planning, execution, and verification modules to achieve closed-loop reasoning from intentions to concrete actions. However, existing agent-based approaches rely on black-box knowledge without theoretical guidance, often yielding unsafe or unverified rules. 
In contrast, our \textbf{BRT} grounds multi-agent collaboration in formal requirements representation by reconstructing intentions from fragmented inputs (bottom-up) and implementing functional intentions and quality concerns(top-down). To the best of our knowledge, our work presents the first method that integrates bidirectional traceability for intent-driven rule completion in IoT.
\vspace{-0.3cm}
\section{Conclusion}

This paper addresses the critical gap between fragmented TAP expressions and high-level user intentions in IoT end-user programming. We propose an intent-driven requirements completion approach that reframes rule completion as a dual process—reconstruct-ing intent from fragmented rules and generating complete rules from that intent—with safety embedded throughout. Thus, we introduce the \textbf{BRT}, a three-layer structured representation that explicitly links rules, intents, and quality concerns, and design a multi-agent collaboration framework that combines the semantic reasoning capabilities of LLMs with structured traceability. Experimental results demonstrate that our method significantly outperforms baselines, improving the rule completion rate by 43\% and reducing logical conflicts by over 21\%. The core contribution of this work lies in driving a paradigm shift: elevating rule completion from mechanical rule-filling to traceable intent reconstruction, shifting the burden of specification from users to the system, and advancing the focus from functional correctness to holistic trustworthiness.

Future work will proceed in three directions: validating dynamic runtime safety through \textit{in situ} studies, generalizing the framework to broader end-user programming paradigms and complex IoT systems, and exploring continuous learning agents to proactively detect and resolve hallucinations during collaboration.

\vspace{-0.3cm}
\section{Data Availability Statement}

The related source codes and datasets are available \url{https://doi.org/10.5281/zenodo.19337953}.


\bibliographystyle{ACM-Reference-Format}
\bibliography{ref}

\end{document}